\documentclass{article}
\usepackage[T1]{fontenc} % add special characters (e.g., umlaute)
\usepackage[utf8]{inputenc} % set utf-8 as default input encoding
\usepackage{ismir,amsmath,cite,url}
\usepackage{graphicx}
\usepackage{color}
\usepackage{pifont}
\usepackage{censor}
\usepackage{booktabs}
\usepackage{amsfonts}

\usepackage{lineno}
\title{Sanidha: A Studio Quality Multi-Modal Dataset for Carnatic Music}

\multauthor
{Venkatakrishnan Vaidyanathapuram Krishnan$^1$ \hspace{1cm} Noel Alben$^1$ \hspace{1cm} Anish Nair$^1$} { \bfseries{Nathaniel Condit-Schultz$^1$}\\
 $^1$ School of Music, Georgia Institute of Technology, USA\\
{\tt\small \{vkrishnan65, noelalben3, anair323, natcs@gatech.edu\}}
}

\def\authorname{V. Vaidyanathapuram Krishnan, N. Alben, A. Nair, and N. Condit-Schultz}

\begin{document}

\maketitle
%
% If you have any questions, please contact the Program Committee (\texttt{ismir\conferenceyear-papers@ismir.net}).
% This template can be downloaded from the ISMIR \conferenceyear\ web site (\texttt{http://ismir\conferenceyear.ismir.net}).

%
\begin{abstract}

Music source separation demixes a piece of music into its individual sound sources (vocals, percussion, melodic instruments, etc.), a task with no simple mathematical solution. It requires deep learning methods involving training on large datasets of isolated music stems. The most commonly available datasets are made from commercial Western music, limiting the models' applications to non-Western genres like Carnatic music. Carnatic music is a live tradition, with the available multi-track recordings containing overlapping sounds and bleeds between the sources.
This poses a challenge to commercially available source separation models like Spleeter and Hybrid Demucs.
In this work, we introduce \textit{Sanidha}, the first open-source novel dataset\footnote{\textit{Sanidha} dataset (Licensed under CC-BY-4.0) is hosted in the server: \texttt{https://ccml.gtcmt.gatech.edu/data/Sanidha}} for Carnatic music, offering studio-quality, multi-track recordings with minimal to no overlap or bleed. 
Along with the audio files, we provide high-definition videos of the artists' performances. 
Additionally, we fine-tuned Spleeter, one of the most commonly used source separation models, on our dataset and observed improved SDR performance compared to fine-tuning on a pre-existing Carnatic multi-track dataset. The outputs of the fine-tuned model with \textit{Sanidha} are evaluated through a listening study.
% This dataset aims to enhance currently available source separation models' applicability to the Indian art form of Carnatic music and push forward research in music information retrieval for the genre.
%Finally, we perceptually evaluate the trained models on the pre-existing dataset.

%% ADD DETAILS ABOUT HOW THE DATA IMPROVES spleeter’s performance.
\end{abstract}
\section{Introduction}\label{sec:introduction}

Carnatic music is a traditional "art music" genre from the Southern part of India. 
Carnatic Music is largely improvised, requiring all musicians to utilize a complex understanding of the melodic and rhythmic structures of the music to improvise coherently.
Carnatic performances generally feature four to five musicians centered around a vocalist in the lead role.
The core instruments are the violin, in both supportive and lead roles; the \textit{mridangam}, a tonal two-sided drum that provides rhythmic support; and the \textit{ghatam}, a clay pot instrument that contributes rhythmic patterns to complement the \textit{mridangam} in a higher frequency range.
% Other common instruments include \textit{kanjira}, \textit{tavil}, \textit{Indian flute}, and the \textit{veena}.
Carnatic Music performances are also accompanied by a \textit{tanpura}, which constantly oscillates the \textit{sa}, the tonic, and either the \textit{pa}, the fifth or sometimes \textit{ma}, the fourth. 
All the instruments are tuned to these frequencies, including the percussion instruments, which, too, have tonal qualities \cite{srinivasamurthy2023getting}. 
This leads to a significant overlap of frequency content, making Carnatic Music source separation almost impossible with simple dictionary learning methods \cite{sebastian2016group}.

Like most traditional music genres, Carnatic Music is performed live \cite{srinivasamurthy2023getting}.
% with little or no use of, or influence from, modern (pop music) production technologies or practices.
Thus, recordings of Carnatic Music lack multi-track isolation, as microphones inevitably capture signals from multiple instruments as well as the audience---these unwanted signals are known in music production as leakage or ``bleed.''
This contrasts with Western pop music, where completely isolated multi-tracks are commonplace, and many source separation datasets are available \cite{manilow2019cutting, rafii2017musdb18, bittner2014medleydb, SignalSep16}. 
The most extensive open-source Music Information Retrieval (MIR) dataset of Indian art music---the \textit{Saraga} dataset \cite{srinivasamurthy2021saraga}---exhibits significant leakage between different audio tracks: For example, the sound of the violin is audible in the vocal track - The bleeding of other sources into other microphones is significant \cite{nuttall2021matrix, plaja2023carnatic, plaja2023repertoire}.

% Source separation has been an active area of music information retrieval research for years.
% Like many areas of music scholarship, this research has been heavily biased towards Western musical styles, and source separation research for non-Western music remains extremely limited \important{(can we cite ANY examples?).}

\subsection{Leakage Problem}

Consider a signal $\mathbf{s}$, noise $\mathbf{n}$, and a mix $\mathbf{x}$, at 0 dB Signal-to-Noise Ratio (SNR): $\mathbf{x} = \mathbf{s} + \mathbf{n}$, where $\mathbf{x}, \mathbf{s}, \mathbf{n} \in \mathbb{R}^d$.
Let $\mathbf{s_t}, \mathbf{n_t} \in \mathbb{R}^d$ such that they represent ground truth signal and noise with bleed.
Assume no microphone sensor noise and no Room Impulse Response (RIR). Then
\begin{equation}
    \mathbf{x = s_t + n_t}
    \label{eq:x}
\end{equation}
\begin{equation}
    \mathbf{s_t} = f(\mathbf{s}, \mathbf{n}) = \alpha\mathbf{s} + \beta\mathbf{n}
    \label{eq:st}
\end{equation}
where $\alpha \in [0,1]$ and $\beta \in [0,1]$, using Eq. \ref{eq:x}, it follows that
\begin{equation}
    \mathbf{n_t} = g(\mathbf{s}, \mathbf{n}) = (1 - \alpha)\mathbf{s} + (1 - \beta)\mathbf{n}
    \label{eq:nt}
\end{equation}
Assume that functions $f$ and $g$ are linear time-invariant functions for all audios.
However, the $\alpha$ and $\beta$ values will vary for different signals in a general unclean dataset.

\begin{figure}
    \centering
    \includegraphics[scale=0.25]{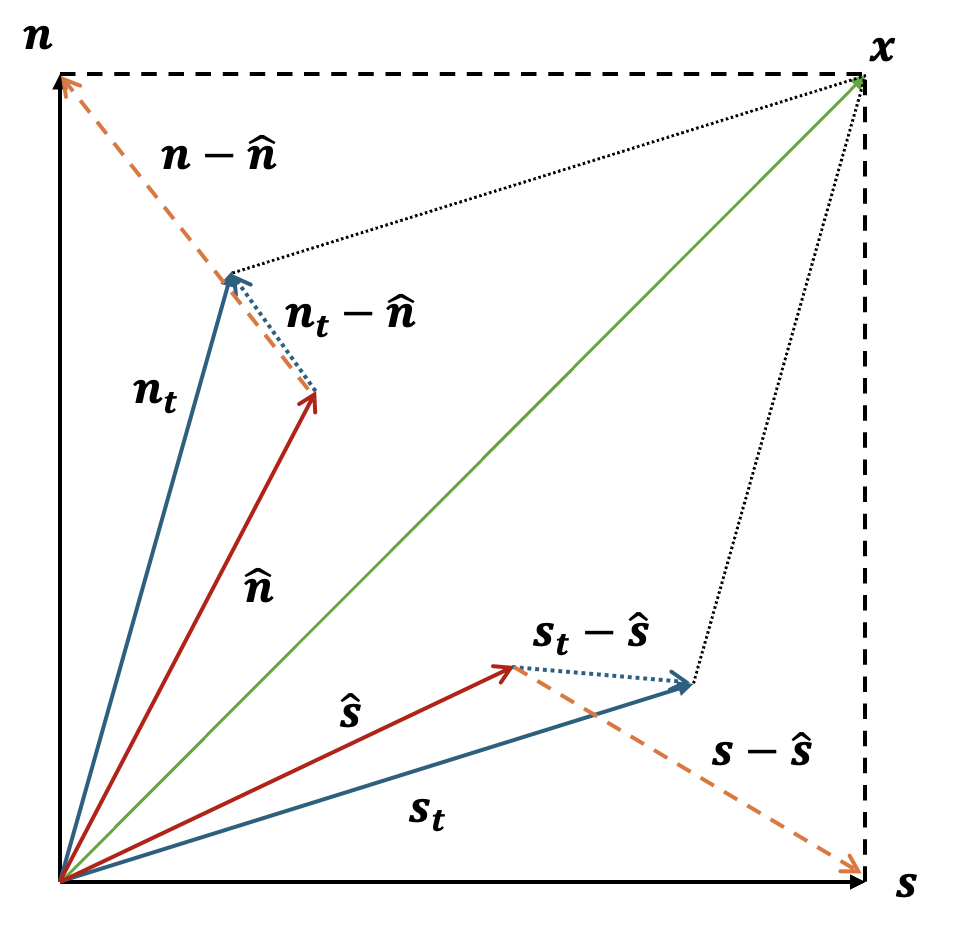}
    \caption{Problem of Poor Ground Truth}
    \label{fig:SDR_poor_gd}
\end{figure}

Let the source separation function trained with $(\mathbf{s_t},\mathbf{n_t})$ as the ground truth be $\mathbf{F}$, such that 
\begin{equation*}
    \mathbf{F}(\mathbf{x}) = (\mathbf{\hat{s}}, \mathbf{\hat{n}})
\end{equation*}
For simplicity, let us assume $\mathbf{\hat{s}}$, $\mathbf{s_t}$ and $\mathbf{n_t}$ lie in the same subspace as $\mathbf{s}$ and $\mathbf{n}$; $\mathbf{s}$ and $\mathbf{n}$ are orthogonal to each other i.e. $\mathbf{s}^T\mathbf{n} = 0$ as seen in Figure \ref{fig:SDR_poor_gd}.

The most common metric used for evaluation and loss in the source separation community is the Signal-to-Distortion Ratio (SDR) and, more recently, the Scale-Invariant version of SDR called SI-SDR \cite{SISDR2019}. 
For simplicity, let's consider using SDR for evaluation since the idea can easily be extended to SI-SDR.
SDR is defined by  \cite{bss_eval_toolbox} for the \texttt{BSS\_eval} toolbox (which is the same as classical SNR) as:

\begin{equation*}
    \text{SDR}_s = 10\log_{10}\bigg(\frac{||\mathbf{s}||}{||\mathbf{s} - \mathbf{\hat{s}}||}\bigg)
\end{equation*}

Given that $f$ and $g$ functions vary for each audio, the SDR formula above is modified for data with bleed as: 

\begin{equation*}
    \text{SDR}_{s, mod} = 10\log_{10}\bigg(\frac{||\mathbf{s_t}||}{||\mathbf{s_t} - \mathbf{\hat{s}}||}\bigg)
\end{equation*}

These objective results from SDR, however good, will never truly represent what the original source must sound like. 
Training on data with a significant bleed will never push the predicted $\mathbf{\hat{s}}$ towards the actual source $\mathbf{s}$, since the loss function will be trained on the modified function dependent on sources with bleeding.

Furthermore, the result will be subpar after incorporating scale invariance \cite{SISDR2019}.
If we calculate the norm of $\mathbf{s_t}$ and $\mathbf{n_t}$, using Eq. \ref{eq:st}, \ref{eq:nt}, and the triangle inequality, we can prove:
% \begin{equation*}
%     ||\mathbf{s_t}|| \leq \alpha||\mathbf{s}|| + \beta||\mathbf{n}||
% \end{equation*}
% \begin{equation*}
%     ||\mathbf{n_t}|| \leq (1-\alpha)||\mathbf{s}|| + (1-\beta)||\mathbf{n}||
% \end{equation*}
\begin{equation}
    ||\mathbf{s_t}|| + ||\mathbf{n_t}|| \leq ||\mathbf{s}|| + ||\mathbf{n}||
    \label{eq:SISDR_problem}
\end{equation}
This means that if we had to calculate the average SI-SDR of the signal and the noise with respect to the sources with bleed, the error would be significant. This error will be large when compared to calculating it with respect to "true" sources, which are inaccessible.
It is also important to note that this was based on the assumption that all were in the same subspace, but that is never true in real scenarios, resulting in increased error.

Hence, the \textit{Saraga} dataset cannot be used as accurate ground truth data for supervised source separation models for both training and especially evaluation, hindering the development of such models for Carnatic Music.
% Using leakage-laden tracks will inhibit the model's capability to learn the target sources properly as well as evaluate them.
As a workaround, some have attempted using source separation models like Spleeter \cite{spleeter2020}, presumably trained on a few or no Carnatic Music examples \cite{plaja2023carnatic}, directly on the vocal multi-track with bleeding for certain MIR tasks \cite{nuttall2021matrix, plaja2023repertoire}.
However, attempts toward source separation for Carnatic using the currently available datasets have been made \cite{sebastian2016group, plaja2023carnatic, dawalatabad2021front}.

The stems obtained for Western Music datasets \cite{manilow2019cutting, rafii2017musdb18, bittner2014medleydb, SignalSep16} are all from studio recordings, recorded separately and mixed, resulting in zero bleeds of other instruments in the multi-tracks.
This allows for evaluation metrics such as SDR, Scale-Invariant SDR \cite{SISDR2019}, Signal-to-Aritfacts Ratio, Signal-to-Interference Ratio (SIR), etc., to be used without problems.
However, there is no such available dataset for Carnatic Music \cite{sebastian2016group, plaja2023carnatic}, and since it is a live tradition, it is impossible to record the artists at separate times.

There have been a lot of datasets for Carnatic Music and Hindustani Music, which provide clean studio-quality data for individual instruments \cite{krishna_subramani_dataset, clayton2021hindustani}.
However, there have been no completely isolated full live concert recordings of studio quality.
To directly address this requirement, we present a new dataset of well-isolated multi-track recordings of Carnatic Music: \textit{Sanidha}.
The \textit{Sanidha} dataset features audio and video recordings of Carnatic musicians playing together in real-time but in total isolation within a modern studio environment. 
\section{Methodology}\label{sec:methodology}

Serra \cite{serra2014corpus} proposed five essential considerations when creating new corpora: purpose, coverage, completeness, quality, and reusability.
These considerations guided the creation of the \textit{Saraga} dataset of Indian art music, \cite{srinivasamurthy2014corpora}, and we have worked to apply the same principles to the construction of \textit{Sanidha}.

% The traditional performance practices of Carnatic music make capturing well-isolated audio a significant challenge.
The isolated tracks for the commercial Western music source separation datasets are often created by the process of overdubbing in the studio.
Carnatic Music must be improvised collectively in real-time, so parts cannot be ``overdubbed" one at a time, thereby posing a significant challenge.
Carnatic musicians listen closely to each others' playing and communicate extensively using visual cues.
In particular, the vocalist often indicates the \textit{taalam} (metric structure) with their hands.
Visual cues are critical during fully improvisational sections like the \textit{kalpana swaram} and \textit{tani avartanam}.
Consequently, the only way to record the music with audio isolation is for each musician to play in separate rooms while maintaining communication through audio and video.

\subsection{Recording Sessions}

We organized five Carnatic music concerts within our recording facility in March of 2024.
Concert sessions lasted 2--3 hours, garnering an average of 1.6 hours of music per concert once silence between pieces was edited out.
To perform these concerts, we recruited fifteen professional Carnatic musicians from Atlanta's thriving Carnatic music scene.
% Some of our senior performers had already significantly contributed to the Carnatic music landscape prior to this endeavor.
% All the volunteering musicians performed with exceptional skill and musicality, especially given the unusual performance environment.
All the artists voluntarily agreed to contribute to the dataset for research purposes, with no compensation\footnote{The concerts were conducted with the approval of the Georgia Tech Institution Review Board (IRB) (ethics board), including two minors who were accompanied by their parents.}.

Our musicians included three male vocalists, two female vocalists, four violinists, and six percussionists.
Two out of five concerts featured a vocalist accompanied by the full set of core Carnatic instruments (violin, \textit{mridangam}, and \textit{ghatam}). 
The other three concerts proceeded without a \textit{ghatam} player---which is not unusual for the style.
Through the efforts of multiple talented musicians, we were able to capture gender diversity in the vocal timbre and a wide array of stylistic and improvisational approaches, which enhances the value of our data to the research community.

\subsection{Recording Facility and Setup}
% Georgia Tech Studio
% Why choose GT Studio?
% What do they provide? - the booths, separators, mixing console etc.
The dataset was recorded in four rooms of the West Village Music Annex, in the Georgia Institute of Technology's campus in Atlanta, Georgia, USA.
These rooms are multi-purpose spaces with large acoustic curtains, which enhanced our ability to control reverberation and maintain adequate isolation.
The four isolated rooms have connection points wired to a single recording control room, including low-impedance, balanced analog audio, and digital video (SDI) connections.
The control room uses a 32-channel digital mixing console to control audio routing and doubles as a multi-channel audio interface for digital audio recording into our Digital Audio Workstation (DAW).
A \textit{tanpura} drone, generated by a \textit{shruti} box or a video from the internet, was also routed to each artist's headphones from the control room.
%We used a variety of \textit{tanpura} drones sounds over the course of the concert series, maximizing variability.
We used the board's onboard reverb, compression, and equalization effects to create custom monitoring mixes sent to their headphones/in-ears, catering to individual artist needs and simulate the live traditional performing scenario of Carnatic Music.

Each artist's performance video was captured using a professional 4K video camcorder. 
The recorded video feed was then delivered through SDI cables from each room to the control room to generate a multi-source mixed feed, allowing us to transmit all four video feeds within a 2x2 grid (Figure \ref{fig:concert3-snapshot}).
Musicians could see the 2x2 feed projected onto a screen in the performance room, allowing them to observe each other at all times.

%\important{(Mention about the video processing and patching)}

% We choose corners for background screens
% Also, we have access to high-level studio equipment. 

\begin{table*}   % title name of the table  
    \centering % centering table 
    \begin{tabular}{c l c c c c c} % creating 10 columns  
        \hline\hline \\ 
         \textbf{Concert} & \textbf{Instruments} & \textbf{Multi-tracks} & \textbf{Front-View Video} & \textbf{Side-View Video} & \textbf{Duration (hr)} & \textbf{Vocals Gender}
        \\ [0.5ex]  
        \hline   
        % Entering 1st row  
        1 & Vocal & 1 & \ding{51} & - & 1.08 & Female \\
         & Violin & 2 & \ding{51} & - & & \\
         & Mridangam & 2 & \ding{51} & - & & \\
         & Ghatam & 2 & \ding{51} & - & & \\
        \hline  % inserts single-line 
        % Entering 2nd row  
        2 & Vocal & 1 & \ding{51} & \ding{51} & 1.63 & Male \\
         & Violin & 2 & \ding{51} & - & & \\
         & Mridangam & 2 & \ding{51} & - & & \\
        \hline  % inserts single-line  
        3 & Vocal & 1 & \ding{51} & \ding{51} & 1.37 & Male \\
         & Violin & 2 & \ding{51} & - & & \\
         & Mridangam & 2 & \ding{51} & - & & \\
         & Ghatam & 2 & \ding{51} & - & & \\
        \hline  % inserts single-line 
        4 & Vocal & 1 & \ding{51} & \ding{51} & 1.97 & Female \\
         & Violin & 2 & \ding{51} & - & & \\
         & Mridangam & 2 & \ding{51} & - & & \\
        \hline  % inserts single-line 
        5 & Vocal & 1 & \ding{51} & \ding{51} & 1.92 & Male \\
         & Violin & 2 & \ding{51} & - & & \\
         & Mridangam & 2 & \ding{51} & - & & \\
        \hline  % inserts single-line 
        \hline\hline  % inserts double-line  
    \end{tabular}  
    \caption{Dataset Details}
    \label{table:data}
\end{table*}

% While performing, each musician wore headphones (in-ear or over-ear) so they could hear the other musicians and their own recorded signal (as needed).

Our musicians had little to no experience performing in a studio setting, isolated from each other, with headphones/in-ears on.
Our efforts were focused on ensuring that the recording sessions were comfortable for the musicians and maintained the ``natural'' performance feeling as much as possible.
Despite our best efforts, our musicians noted specific challenges performing within the constraints of the setup and sometimes felt that it slightly affected the quality of their performance.

Though our audio-monitoring setup achieved close to zero latency, we found that our video-monitoring setup lagged by about 50 ms, possibly due to the converters used to transmit the video feed to the projectors. 
This made it extremely difficult for the artists to coordinate with each other, since they could not follow the \textit{taalam} or beat given by the vocalist. 
To overcome this problem, we used a \textit{proxy-taalam} setup.
One of our team members would sit in front of each artist (except the vocalist) and provide the visual \textit{taalam} cue by focusing on just the audio feed from the vocalist.
This setup was most helpful for our percussion artists; even the violinists appreciated it during the improvised \textit{kalpana swaram} sections. 
The \textit{proxy-taalam} setup allowed the musicians to play in time with each other, react to the cues from the vocalist similar to a live concert setup, and make the improvisational sections of Carnatic Music - \textit{tani avartanam} and \textit{kalpana swaram} sections possible.

We also identified a potential issue much later when we observed that some artists partially removed one side of their headphones in the middle of their performance.
In some cases, artists required loud headphone output.
This resulted in slight bleeding of the headphone output to the performer's microphone. 
To combat this, we shifted the monitoring system to in-ear monitors exclusively for all further concerts, which nullifies possible bleeding from headphones.

\subsection{Audio Data}

For each concert, we recorded six (excluding \textit{ghatam}) or eight (full group) separate unprocessed audio tracks.
Vocalists were recorded using a single microphone;
the other instruments were recorded using two microphones each. 
We captured the violin and \textit{mridangam} in a standard stereo (left-right) image. The \textit{ghatam} recording setup used two microphones as well.
A line-in track was used to record the \textit{tanpura} drone.

In total, we have nearly eight hours of recorded music, across the five concert sessions.
The recorded audio is in WAV format, with CD-standard sampling rate of $44.1$ kHz and a bit depth of 16 bits.
Table \ref{table:data} displays all the individual concert durations.

\subsubsection{Microphones}

% Microphones used for each instrument

For each concert, we used different combinations of microphones, maximizing the sonic variety of the data.
The choices of microphones were professional, studio-grade condenser microphones with cardioid polar pickup patterns, with each instrument requiring matched pairs of identical microphones.
The use of high-fidelity condenser microphones contrasts with the dynamic microphones used commonly in traditional Carnatic music concerts.
However, capturing the highest fidelity audio will produce the most broadly usable data.
A series of non-linear operations can be performed at the post-processing stage to alter high-fidelity signals to sound more like dynamic microphones.
The details of the microphones used for each instrument are stored in a JSON file located within respective concert folders.

For our first concert, the vocal microphone was placed close to the vocalist's mouth.
We realized that this position obstructed the video of the performer's face. 
For all subsequent concerts, we corrected this by placing microphones closer to chest level, pointing upwards towards the mouth, ensuring an obstruction-free video. 

We placed microphones for the violin and \textit{mridangam} on either side of the artist, at a distance of approximately 50 cm.
This positioning ensured microphone stability, kept the video feed unobstructed, and highlighted each instrumentalist's gestures and hand movements.
As the \textit{ghatam} is a relatively quiet instrument, we placed the first microphone as close as possible to the playing surface.
The second was pointed toward the opening of the \textit{ghatam} at a distance of $\approx$ 30 cm.
This can be seen in Figure \ref{fig:concert3-snapshot}.

\vspace{-5mm}

\subsection{Video Data}

Performance video data for Carnatic is significantly limited compared to Hindustani music. 
The access to video data has given rise to a significant interest in the multi-modal analysis of Hindustani music among the MIR community \cite{paschalidou2023multimodal, clayton2022raga, clayton2021hindustani, kelkar2015applications}. Our motivation to include video recordings with our dataset is to promote multi-modal research endeavors for Carnatic music.

\begin{figure}
    \centering
    \includegraphics[scale=0.1]{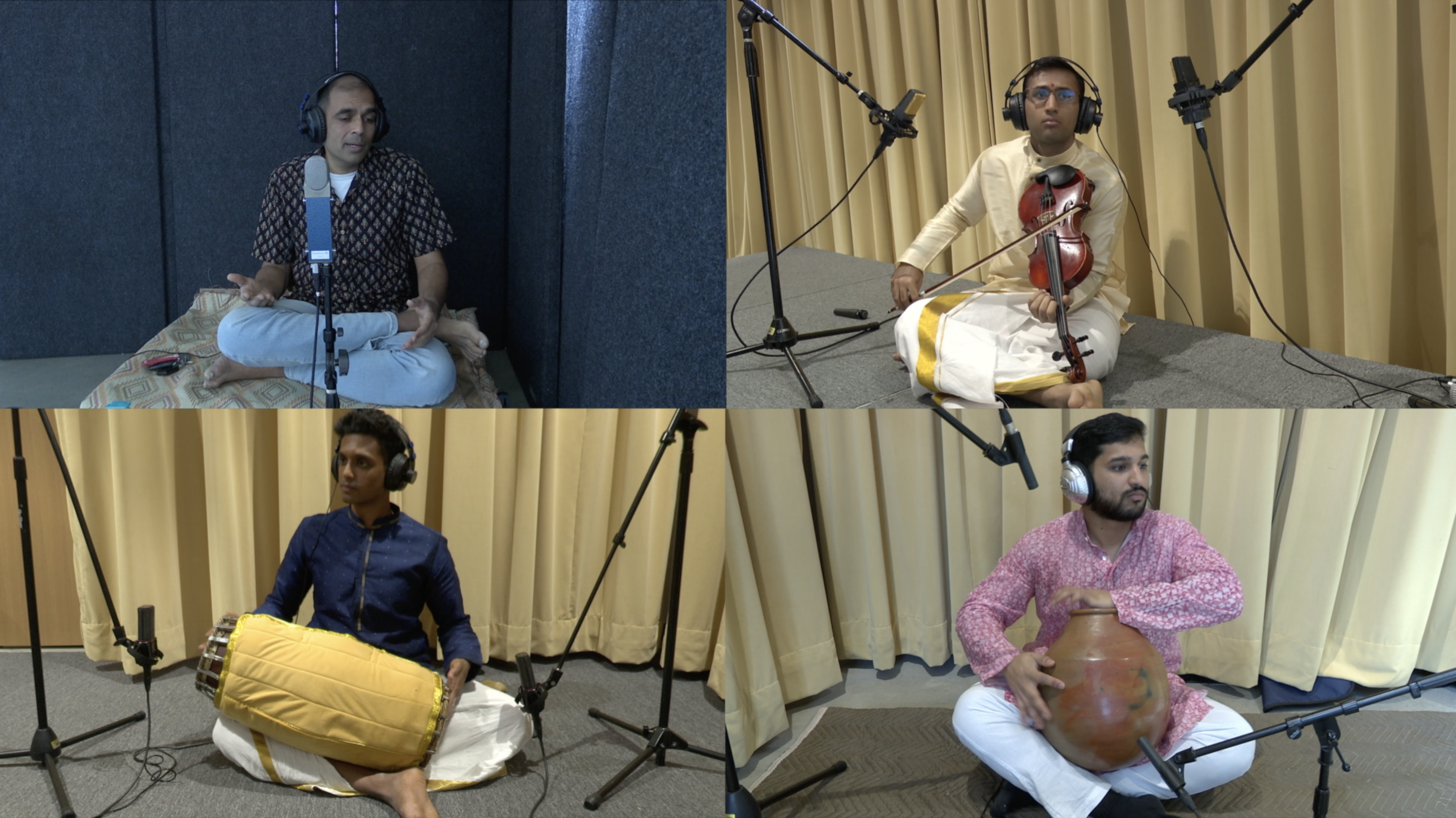}
    \caption{Snapshot of the Front-view videos of Concert 3}
    \label{fig:concert3-snapshot}
\end{figure}

All of our videos are recorded at 29.97 FPS in 1080p. 
The snapshot of the front view videos of each instrumentalist can be seen in Figure \ref{fig:concert3-snapshot}.
The lighting for all the videos takes advantage of the many light sources available in the multi-purpose recording rooms.

For each concert, we successfully captured the front-view videos of every musician and included an additional side view of the vocalist. This combination is a first for a dataset of this kind.

The framing of the front-view videos is
similar to the stills used in \cite{clayton2022raga}.
To ensure a solid background, we placed solid black sound panels behind the vocalists and solid yellow curtains behind the other artists, as seen in Figure \ref{fig:concert3-snapshot}.

\subsection{Supplementary Information}

\subsubsection{Metadata}

To fulfill Serra's completeness criteria, we collected annotations and metadata similar to \textit{Saraga} \cite{srinivasamurthy2021saraga}.
This metadata is stored in separate JSON files for each song performed during the concerts. The metadata includes the composition name, original composer, and the performers' names and roles. 
We also include relevant music-theory information regarding the compositions, mentioning the rāgam, tālam, and song form.

\subsubsection{Section Annotations}

The song form is encoded as audio timestamps indicating the start and end of each major musical section for every song: 
the key sections are the \textit{aalapana}, \textit{pallavi}, \textit{anupallavi}, \textit{muktayi swaram}, \textit{charanam}, \textit{cittai swaram}, \textit{kalpana swaram} and \textit{neraval}.
The performing musicians were consulted to review all of the metadata.

\subsubsection{Pitch Annotations}

Carnatic music contains two melody sources: the lead vocals and the violin, which complements the vocals. 
Since we have clean vocals and violin data, the Melodia algorithm proposed by Salamon and Gòmez \cite{salamon2012melodia} was used to extract pitch (F0) contours for these two parts.
The pitch tracks are stored in a two-column format, with the time stamps in the first column and the pitch values in the second.

\subsubsection{Tonic Annotations}

Obtaining the tonic frequency is relatively easy since we have a clean \textit{tanpura} source within our multi-track data.
We followed a similar approach used by Gulati et al. \cite{gulati2012tonic} and used Melodia \cite{salamon2012melodia} on the \textit{tanpura} multi-track directly for the tonal feature extraction.
The tonic does not change within a concert; hence, we included a single tonic file, which stores the tonic value in Hertz, inside each concert folder instead of having one for every song.

% Well-documented file available at GitHub
% Along with code
\begin{table*}[t]
    \centering
    \begin{tabular}{|c|c|c||c|c|c|c|c|c|c||c|c|}  % repeats {c|} 14 times
    \hline
         \multicolumn{2}{|c}{}& & \multicolumn{4}{|c}{\textbf{\textit{Sanidha}} - Objective Evaluation} & \multicolumn{2}{|c|}{\textbf{\textit{Saraga}} - Perceptual Evaluation} \\ \hline \hline
        \textbf{Models} & \textbf{Hours} & \textbf{Source} &  \textbf{SDR} & \textbf{SIR} & \textbf{SAR} & \textbf{SI-SDR} &  \textbf{Isolation} & \textbf{Audio Quality} \\ \hline \hline
        \textit{Saraga} & 12.37 & Vocals & 7.66 & 17.05 & 8.02 & 6.65 & 0.596 & 0.627 \\
        & & Accomp. & 7.68 & 13.65 & 8.84 & 7.29 & 0.546 & \textbf{0.532} \\ \cline{3-9}
        & & \textbf{Average} & 7.67 & 15.35 & 8.43 & 6.97 & 0.564 & \textbf{0.580} \\\hline \hline
        
        \textit{Sanidha} & 13.21 & Vocals & \textbf{7.86} & \textbf{17.38} & \textbf{8.26} & \textbf{6.93} & 0.598 & \textbf{0.635} \\
         & & Accomp. & \textbf{7.87} & 13.96 & \textbf{8.99} & \textbf{7.52} & 0.541 & 0.507 \\ \cline{3-9}
         & & \textbf{Average} & \textbf{7.87} & \textbf{15.67} & \textbf{8.63} & \textbf{7.22} & 0.570 & 0.572 \\ \hline \hline
         
        Mix & 12.37 + 13.21 & Vocals & 7.63 & 16.88 & 8.00 & 6.62 & \textbf{0.605} & 0.621 \\
         & & Accomp. & 7.65 & \textbf{13.99} & 8.73 & 7.25 & \textbf{0.561} & 0.525 \\ \cline{3-9}
         & & \textbf{Average} & 7.64 & 15.44 & 8.36 & 6.93 & \textbf{0.583} & 0.573 \\ \hline
    \end{tabular}
    \caption{Results}
    \label{tab:results}
\end{table*}

\section{Experiments}

The experiments aim to cover the Coverage and Quality principles \cite{serra2014corpus} introduced in Section \ref{sec:methodology} and demonstrate the value and usability of our new dataset with a simple source separation experiment.

% As proved in Section \ref{sec:introduction}, we shall also show experimentally that the objective results of the models match with the perceptual test results on the Saraga Dataset \cite{srinivasamurthy2021saraga}, which represents real-life Carnatic concerts.

% We also change the speed (pitch+time) of all the audios by ±4\% for augmentation by resampling techniques. 
% This ensures naturality in the sound and transients instead of plain time-stretch or pitch shifting.
% One important note is that we do not clear the \textit{taalam} given by the vocalist in the empty vocal sections.

% Our goal is to compare three source separation models trained \textit{Sanidha}, \textit{Saraga} and a training procedure involving both. 
It is important to note that our aim in this work is to demonstrate our data's potential through these preliminary experiments and not benchmark performance against the state-of-the-art results for source separation of Carnatic Music.

\subsection{Experiment Setup}

We ran a simple two-stem source separation fine-tuning experiment on \textit{Sanidha} and \textit{Saraga} datasets using the Spleeter model \cite{spleeter2020}.
Two-stem Spleeter training requires the vocals, accompaniment, and mix audios.
We fine-tuned the pre-trained model using three different approaches: (1) using the \textit{Sanidha} dataset, (2) using the \textit{Saraga} dataset, (3) using curriculum training \cite{bengio2009curriculum, wang2021survey} by partly fine-tuning the model with \textit{Saraga}, and then fine-tuning it further with the \textit{Sanidha} dataset. The curriculum training strategy presents the data to the model in a meaningful order to learn better.
% The model, which is trained on a mix of both, does not contain a merged dataset but follows a curriculum learning \cite{bengio2009curriculum, wang2021survey} methodology, where we present the data to the model in a meaningful order to learn better.
Using these three models will help us evaluate the potential of our data and its performance when combined with other Carnatic Datasets, in this case, \textit{Saraga}.

Since \textit{Sanidha} has fewer concerts than \textit{Saraga}, the major problem which could arise, is the possibility of overfitting.
To potentially avoid this, the third model is fine-tuned on \textit{Saraga} for 225K steps (90\% of the total steps), while the rest 10\% is finetuned on \textit{Sanidha} for 25K steps.

\subsection{\textit{Sanidha} Data Preparation}
\label{subsec:data_prep}
\textit{Sanidha's} audio data is of high quality as it was recorded in isolated spaces using condenser microphones with almost no bleed.
Therefore, just linearly adding the signals to prepare mixes for training \cite{opensourceseparation:book} will not be representative of the traditional Carnatic Concerts.
To prevent this, we chose ten concerts from the \textit{Saraga} dataset and used them as reference tracks to create two unique mixes for each of the five \textit{Sanidha} concerts.
Eight out of these ten \textit{Saraga} tracks are used as references for processing the training set and the remaining two are used for validation. 
The multiple mixes allow us to obtain twice the original amount of data.
This can be considered as data augmentation since we have limited clean data.
Our goal was to match the number of hours of training data used on the models individually trained on \textit{Sanidha} and \textit{Saraga} respectively, to make a fair comparison. 
The \textit{Sanidha} training set makes up a total of 13.21 hours of audio data, and the validation set is 2.14 hours.

The critical mixing strategies for vocals and accompaniment include a combination of multiple non-linear and some time-varying operations - (1) Adding distortion, (2) Adding white noise, (3) Processing the stems through a digital amplifier plus cabinet models, (4) Heavy compression, (5) Adding reverb, (6) Attenuating the body of the instruments and vocals, and (7) High-cut filtering.
Each of these operations is performed in varied amounts to match the sonic features of the reference tracks. The aim is to mix the tracks to emulate a real live concert while maintaining the isolated ground-truth audio.

The processed vocals ($\mathbf{v}$) and the processed accompaniment ($\mathbf{a}$) audios are linearly added at 0 dB SNR to create the mixture file ($\mathbf{m} = \mathbf{v} + \mathbf{a}$) for training.
For the SNR computation, we consider the signal to be the vocals and the noise to be the accompaniment.

% \begin{equation*}
%     \mathbf{m} = \mathbf{v} + \mathbf{a}
% \end{equation*}

% One of the mixes for each concert is part of the dataset itself.
The third \textit{Sanidha} concert was chosen for the validation set, as it has all of the typical instruments, including the \textit{ghatam}. The rest of the concerts used in the training set maintain a good distribution of vocalist's gender and vocal timbre, as seen in Table \ref{table:data}.
We made two unique mixes for each song in the validation set, which totaled to 2.14 hours of mixture audio data.
\vspace{-2.6mm}

\subsection{\textit{Saraga} Data Preparation}
Seven out of the eight references \textit{Saraga} concerts described in Section \ref{subsec:data_prep} make up the training set for the \textit{Saraga}-trained model.
As \textit{Saraga} consists of live multi-track recordings from Carnatic concerts, the accompaniment audio is created by linearly adding all the multi-track audios except the vocals.
For the validation set of the \textit{Saraga}-trained model, we selected the same reference concerts from \textit{Saraga} that were used to create the mixes for the validation set in Section \ref{subsec:data_prep}.

% The two reference concerts used for the validation set in the \textit{Sanidha}-trained model is used as validation for the \textit{Saraga}-trained model.
The ground truth multi-tracks used have an inherent bleed in them \cite{nuttall2021matrix, plaja2023carnatic, plaja2023repertoire}, as described in Section \ref{sec:introduction}.
The purpose of using a noisy validation set from \textit{Saraga} is to evaluate the model purely trained on \textit{Saraga}, assuming the \textit{Sanidha} dataset never existed.
However, the metrics obtained in Table \ref{tab:results} are on the validation set used for \textit{Sanidha} training.
The total training duration comes to 12.37 hours. 
The remaining unused concerts in \textit{Saraga} are used for the perceptual tests.

\section{Evaluation}

\subsection{Objective evaluation}

We compute the SDR, SIR, SAR, and also the SI-SDR of each of the models for the \textit{Sanidha} validation set.
Table \ref{tab:results} displays all the results.

We can see that the \textit{Sanidha}-trained model has outperformed, however marginally, in all the objective metrics for vocals and the accompaniment separation.
The improvement is only slight, perhaps because \textit{Sanidha} was only trained on four concerts, while \textit{Saraga} is trained on seven (which would mean seven unique vocalists as compared to four), even if the training hours are comparable.
Also, the curriculum training technique performs almost similar to the \textit{Saraga}-trained model.

\begin{figure}
    \centering
    \includegraphics[scale=0.25]{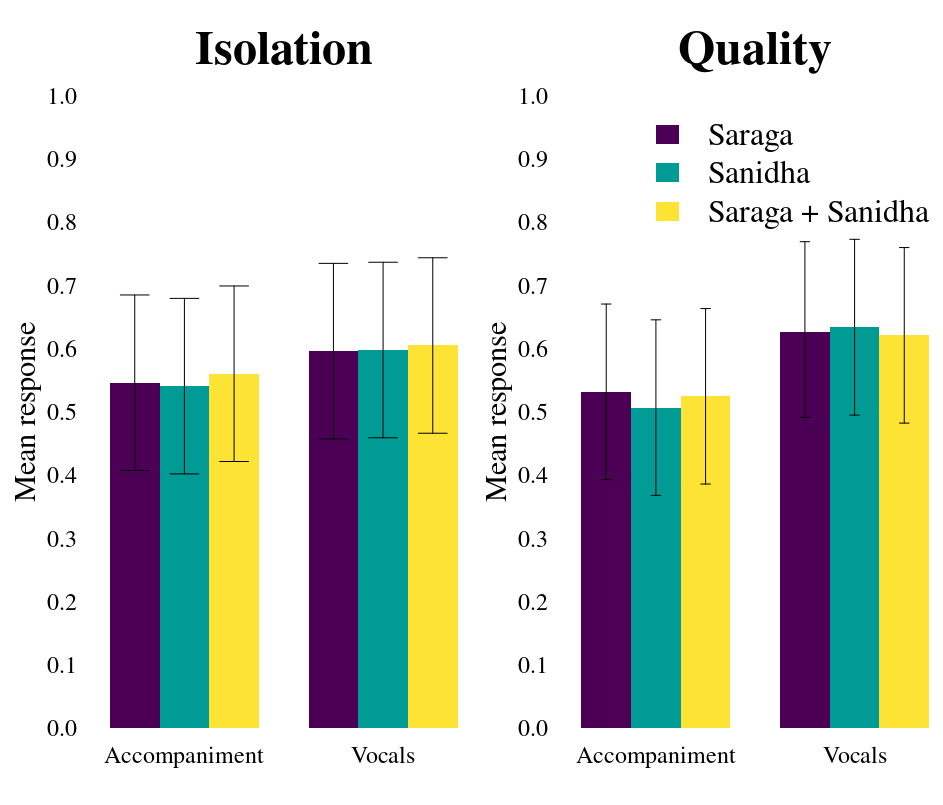}
    \caption{Mean participant responses across twelve conditions, with 95\% confidence limits.}
    \label{fig:anova}
\end{figure}

\subsection{Subjective evaluation}

We conducted a listening study to evaluate the three source separation models and assess their perceptual effectiveness in isolating vocals and accompaniments in Carnatic Music. 

The audio stimuli were selected after we randomly sampled four ten-second excerpts from four different \textit{Saraga} recordings;
If the randomly selected excerpt did not contain the three key instruments in Carnatic Music (vocals, violin, and the \textit{mridangam}), we sampled again until an appropriate excerpt was identified.
This iterative process ensured that our evaluation remains focused on relevant audio features while maintaining the unbiased nature of the sample selection.

In the listening study---approved by the Georgia Tech ethics board---fourteen participants listened to processed versions of our selected excerpts.
The survey was conducted in a manner similar to the MUSHRA framework \cite{schoeffler2018webmushra}. 
All the participants responded to twelve questions for each excerpt, which focused on vocal isolation, vocal audio quality, accompaniment isolation, and accompaniment audio quality for the three models.
This resulted in 48 questions per participant.
These terms have been commonly used in subjective testing of source separation models \cite{plaja2023carnatic, defossez2021hybrid}.
We used a slider-based metric for the evaluation, ranging from zero to one.
Isolation and quality were explained with examples before the start of the survey and also presented as a reference for each question.

Average slider responses for the twelve conditions are shown in Figure \ref{fig:anova} and in Table \ref{tab:results}.
We conducted a mixed-effects ANOVA on the data, with the participant and excerpt as random intercepts and the three variables (response type, target source, and model) as fixed effects.
No effect was statistically significant, except for the target source (voice vs accompaniment), where participants tended to rate vocals higher in general ($\chi^2(8) = 45.97$, $p < .05$).
This behavior is very similar to the objective results as well.

\section{Conclusion and Future Work}
Although fine-tuning spleeter using \textit{Sanidha} did not result in a significant source separation improvement, we cannot discount the importance of the availability of clean target sources for source separation. This is a clear distinction and advantage that our dataset collection methodology has over the existing \textit{Saraga}. We can now use common metrics for source separation evaluation with a good degree of accuracy using our dataset, which was not possible with the existing \textit{Saraga} dataset.
Given the inherent challenges, our introduction of the \textit{Sanidha} dataset marks a significant advancement in this domain.
This novel dataset also presents an avenue for solving a multitude of other MIR and multi-modal tasks in Carnatic Music. 

We will soon expand our dataset and invite more musicians to conduct concerts using our methodology. With the resources at hand, we aim to promote computational analysis for Indian Art music and pave the path towards more accessible research resources within the community.

\section{Acknowledgements}
We extend our gratitude to all the artists whose contributions have been pivotal to the creation of this dataset:
\begin{enumerate}
    \item \textbf{Concert 1:} 
    Amita Krishnan (Vocals), Sudharshan Prasanna (Violin), Vajraang Kamat (Mridangam), Tejas Veedhulur (Ghatam)

    \item \textbf{Concert 2:} 
    Salem Shriram (Vocals), Vishal Sowmyan (Violin), Anirudhah Narayanan (Mridangam)

    \item \textbf{Concert 3:} 
    Prashant Krishnamoorthy (Vocals), Nivik Sanjay Bharadwaj (Violin), Arvind Narayan (Mridangam), Vajraang Kamat (Ghatam)

    \item \textbf{Concert 4:} 
    Anjana Nagaraja (Vocals), Pranavi Srinivasa (Violin), Arvind Narayan (Mridangam)

    \item \textbf{Concert 5:} 
    Prasanna Soundararajan (Vocals), Vishal Sowmyan (Violin), Santosh Chandru (Mridangam)
\end{enumerate}

We also express our gratitude to --
\begin{itemize}
    \item Farshad Jafari from the Georgia Institute of Technology for his assistance in hosting the dataset.
    \item MTG, Barcelona for their time and guidance for the source separation experiments.
\end{itemize}

% \section{Ethics Statement}

% This optional section can be used to provide additional ethical considerations related to your paper, and can be included
% both at submission time and in your camera ready version. See the Call for Papers for details. This section does \textit{not} 
% count towards the page limit for scientific content.

% % For bibtex users:
\bibliography{ISMIRtemplate}

\begin{thebibliography}{10}
\providecommand{\url}[1]{#1}
\csname url@samestyle\endcsname
\providecommand{\newblock}{\relax}
\providecommand{\bibinfo}[2]{#2}
\providecommand{\BIBentrySTDinterwordspacing}{\spaceskip=0pt\relax}
\providecommand{\BIBentryALTinterwordstretchfactor}{4}
\providecommand{\BIBentryALTinterwordspacing}{\spaceskip=\fontdimen2\font plus
\BIBentryALTinterwordstretchfactor\fontdimen3\font minus \fontdimen4\font\relax}
\providecommand{\BIBforeignlanguage}[2]{{%
\expandafter\ifx\csname l@#1\endcsname\relax
\typeout{** WARNING: IEEEtran.bst: No hyphenation pattern has been}%
\typeout{** loaded for the language `#1'. Using the pattern for}%
\typeout{** the default language instead.}%
\else
\language=\csname l@#1\endcsname
\fi
#2}}
\providecommand{\BIBdecl}{\relax}
\BIBdecl

\bibitem{srinivasamurthy2023getting}
A.~Srinivasamurthy, S.~Gulati, R.~Caro~Repetto, and X.~Serra, ``Getting started on computational musicology and music information research: an indian art music perspective,'' \emph{Rao P, Murthy HA, Prasann SRM, editors. Indian art music: a computational perspective.[New Delhi]: Scheme for Promotion Academic and Research Collaboration, 2023. p. 3-38.}, 2023.

\bibitem{sebastian2016group}
J.~Sebastian and H.~A. Murthy, ``Group delay based music source separation using deep recurrent neural networks,'' in \emph{2016 International Conference on Signal Processing and Communications (SPCOM)}.\hskip 1em plus 0.5em minus 0.4em\relax IEEE, 2016, pp. 1--5.

\bibitem{manilow2019cutting}
E.~Manilow, G.~Wichern, P.~Seetharaman, and J.~Le~Roux, ``Cutting music source separation some {Slakh}: A dataset to study the impact of training data quality and quantity,'' in \emph{Proc. IEEE Workshop on Applications of Signal Processing to Audio and Acoustics (WASPAA)}.\hskip 1em plus 0.5em minus 0.4em\relax IEEE, 2019.

\bibitem{rafii2017musdb18}
Z.~Rafii, A.~Liutkus, F.-R. St{\"o}ter, S.~I. Mimilakis, and R.~Bittner, ``Musdb18-a corpus for music separation,'' 2017.

\bibitem{bittner2014medleydb}
R.~M. Bittner, J.~Salamon, M.~Tierney, M.~Mauch, C.~Cannam, and J.~P. Bello, ``Medleydb: A multitrack dataset for annotation-intensive mir research.'' in \emph{ISMIR}, vol.~14, 2014, pp. 155--160.

\bibitem{SignalSep16}
A.~Liutkus, F.-R. St{\"o}ter, Z.~Rafii, D.~Kitamura, B.~Rivet, N.~Ito, N.~Ono, and J.~Fontecave, ``The 2016 signal separation evaluation campaign,'' in \emph{Latent Variable Analysis and Signal Separation - 12th International Conference, {LVA/ICA} 2015, Liberec, Czech Republic, August 25-28, 2015, Proceedings}, P.~Tichavsk{\'y}, M.~Babaie-Zadeh, O.~J. Michel, and N.~Thirion-Moreau, Eds.\hskip 1em plus 0.5em minus 0.4em\relax Cham: Springer International Publishing, 2017, pp. 323--332.

\bibitem{srinivasamurthy2021saraga}
A.~Srinivasamurthy, S.~Gulati, R.~C. Repetto, and X.~Serra, ``Saraga: Open datasets for research on indian art music,'' \emph{Empirical Musicology Review}, vol.~16, no.~1, pp. 85--98, 2021.

\bibitem{nuttall2021matrix}
T.~Nuttall, G.~Plaja-Roglans, L.~Pearson, and X.~Serra, ``The matrix profile for motif discovery in audio-an example application in carnatic music,'' in \emph{International Symposium on Computer Music Multidisciplinary Research}.\hskip 1em plus 0.5em minus 0.4em\relax Springer, 2021, pp. 228--237.

\bibitem{plaja2023carnatic}
G.~Plaja-Roglans, M.~Miron, A.~Shankar, and X.~Serra, ``Carnatic singing voice separation using cold diffusion on training data with bleeding,'' 2023.

\bibitem{plaja2023repertoire}
G.~Plaja-Roglans, T.~Nuttall, L.~Pearson, X.~Serra, and M.~Miron, ``Repertoire-specific vocal pitch data generation for improved melodic analysis of carnatic music,'' \emph{Transactions of the International Society for Music Information Retrieval}, Jun 2023.

\bibitem{SISDR2019}
J.~L. Roux, S.~Wisdom, H.~Erdogan, and J.~R. Hershey, ``Sdr – half-baked or well done?'' in \emph{ICASSP 2019 - 2019 IEEE International Conference on Acoustics, Speech and Signal Processing (ICASSP)}, 2019, pp. 626--630.

\bibitem{bss_eval_toolbox}
E.~Vincent, R.~Gribonval, and C.~Fevotte, ``Performance measurement in blind audio source separation,'' \emph{IEEE Transactions on Audio, Speech, and Language Processing}, vol.~14, no.~4, pp. 1462--1469, 2006.

\bibitem{spleeter2020}
\BIBentryALTinterwordspacing
R.~Hennequin, A.~Khlif, F.~Voituret, and M.~Moussallam, ``Spleeter: a fast and efficient music source separation tool with pre-trained models,'' \emph{Journal of Open Source Software}, vol.~5, no.~50, p. 2154, 2020, deezer Research. [Online]. Available: \url{https://doi.org/10.21105/joss.02154}
\BIBentrySTDinterwordspacing

\bibitem{dawalatabad2021front}
N.~Dawalatabad, J.~Sebastian, J.~Kuriakose, C.~C. Sekhar, S.~Narayanan, and H.~A. Murthy, ``Front-end diarization for percussion separation in taniavartanam of carnatic music concerts,'' \emph{arXiv preprint arXiv:2103.03215}, 2021.

\bibitem{krishna_subramani_dataset}
\BIBentryALTinterwordspacing
K.~Subramani and P.~Rao, ``Carnatic violin dataset,'' 2020. [Online]. Available: \url{https://doi.org/10.5281/zenodo.3940330}
\BIBentrySTDinterwordspacing

\bibitem{clayton2021hindustani}
M.~Clayton, J.~Li, A.~R. Clarke, M.~Weinzierl, L.~Leante, and S.~Tarsitani, ``Hindustani raga and singer classification using pose estimation,'' 2021.

\bibitem{serra2014corpus}
X.~Serra, ``Creating research corpora for the computational study of music: the case of the compmusic project,'' in \emph{AES 53rd International Conference: Semantic Audio; 2014 Jan 27-29; London, UK. New York: Audio Engineering Society; 2014. Article number 1-1 [9 p.].}\hskip 1em plus 0.5em minus 0.4em\relax Audio Engineering Society, 2014.

\bibitem{srinivasamurthy2014corpora}
A.~Srinivasamurthy, G.~K. Koduri, S.~Gulati, V.~Ishwar, and X.~Serra, ``Corpora for music information research in indian art music,'' in \emph{Georgaki A, Kouroupetroglou G, eds. Proceedings of the 2014 International Computer Music Conference, ICMC/SMC; 2014 Sept 14-20; Athens, Greece.[Michigan]: Michigan Publishing; 2014.}\hskip 1em plus 0.5em minus 0.4em\relax Michigan Publishing, 2014.

\bibitem{paschalidou2023multimodal}
S.~Paschalidou and I.~Miliaresi, ``Multimodal deep learning architecture for hindustani raga classification,'' \emph{Sensors \& Transducers}, vol. 260, no.~2, pp. 77--86, 2023.

\bibitem{clayton2022raga}
M.~Clayton, P.~Rao, N.~N. Shikarpur, S.~Roychowdhury, and J.~Li, ``Raga classification from vocal performances using multimodal analysis.'' in \emph{ISMIR}, 2022, pp. 283--290.

\bibitem{kelkar2015applications}
T.~Kelkar, ``Applications of gesture and spatial cognition in hindustani vocal music,'' Ph.D. dissertation, International Institute of Information Technology Hyderabad, India, 2015.

\bibitem{salamon2012melodia}
J.~Salamon and E.~G{\'o}mez, ``Melody extraction from polyphonic music signals using pitch contour characteristics,'' \emph{IEEE transactions on audio, speech, and language processing}, vol.~20, no.~6, pp. 1759--1770, 2012.

\bibitem{gulati2012tonic}
S.~Gulati, ``A tonic identification approach for indian art music,'' \emph{Unpublished master’s thesis). Universitat Pompeu Fabra, Barcelona}, 2012.

\bibitem{bengio2009curriculum}
Y.~Bengio, J.~Louradour, R.~Collobert, and J.~Weston, ``Curriculum learning,'' in \emph{Proceedings of the 26th annual international conference on machine learning}, 2009, pp. 41--48.

\bibitem{wang2021survey}
X.~Wang, Y.~Chen, and W.~Zhu, ``A survey on curriculum learning,'' \emph{IEEE transactions on pattern analysis and machine intelligence}, vol.~44, no.~9, pp. 4555--4576, 2021.

\bibitem{opensourceseparation:book}
\BIBentryALTinterwordspacing
E.~Manilow, P.~Seetharman, and J.~Salamon, \emph{Open Source Tools \& Data for Music Source Separation}.\hskip 1em plus 0.5em minus 0.4em\relax https://source-separation.github.io/tutorial, 2020. [Online]. Available: \url{https://source-separation.github.io/tutorial}
\BIBentrySTDinterwordspacing

\bibitem{schoeffler2018webmushra}
M.~Schoeffler, S.~Bartoschek, F.-R. St{\"o}ter, M.~Roess, S.~Westphal, B.~Edler, and J.~Herre, ``webmushra—a comprehensive framework for web-based listening tests,'' \emph{Journal of Open Research Software}, vol.~6, no.~1, p.~8, 2018.

\bibitem{defossez2021hybrid}
A.~D{\'e}fossez, ``Hybrid spectrogram and waveform source separation,'' \emph{arXiv preprint arXiv:2111.03600}, 2021.

\end{thebibliography}

% For non bibtex users:
%\begin{thebibliography}{citations}
% \bibitem{Author:17}
% E.~Author and B.~Authour, ``The title of the conference paper,'' in {\em Proc.
% of the Int. Society for Music Information Retrieval Conf.}, (Suzhou, China),
% pp.~111--117, 2017.
%
% \bibitem{Someone:10}
% A.~Someone, B.~Someone, and C.~Someone, ``The title of the journal paper,''
%  {\em Journal of New Music Research}, vol.~A, pp.~111--222, September 2010.
%
% \bibitem{Person:20}
% O.~Person, {\em Title of the Book}.
% \newblock Montr\'{e}al, Canada: McGill-Queen's University Press, 2021.
%
% \bibitem{Person:09}
% F.~Person and S.~Person, ``Title of a chapter this book,'' in {\em A Book
% Containing Delightful Chapters} (A.~G. Editor, ed.), pp.~58--102, Tokyo,
% Japan: The Publisher, 2009.
%
%
%\end{thebibliography}

\end{document}